\documentclass[pre,aps,onecolumn,superscriptaddress,nofootinbib]{revtex4-1}
\usepackage{amsmath, amsthm, amssymb,float}
\usepackage{amsfonts}
\usepackage{graphicx}
\usepackage{dcolumn}
\usepackage{bm}
\usepackage{relsize}
\usepackage{textcomp}
\usepackage[normalem]{ulem}
\usepackage[titletoc,title]{appendix}

\usepackage[colorlinks=true, urlcolor=blue, anchorcolor=blue, citecolor=blue,filecolor=blue,linkcolor=blue,menucolor=blue,pagecolor=blue]{hyperref}

\usepackage{epsfig}
\usepackage{textcomp}
\usepackage{epstopdf}

\begin{document}
\title{Large deviations of a long-time average in the Ehrenfest Urn Model}
\author{Baruch Meerson}
\email{meerson@mail.huji.ac.il}
\affiliation{Racah Institute of Physics, Hebrew University of
Jerusalem, Jerusalem 91904, Israel}
\author{Pini Zilber}
\affiliation{Racah Institute of Physics, Hebrew University of
Jerusalem, Jerusalem 91904, Israel}

\begin{abstract}
Since its inception in 1907, the Ehrenfest urn model (EUM) has served as a test bed of key concepts of
statistical mechanics. Here we employ this model to study large deviations of a time-additive quantity.  We consider two continuous-time versions of the EUM with $K$ urns and $N$ balls: without and with interactions between the balls in the same urn.  We evaluate the probability distribution $\mathcal{P}_T(\bar{n}= a N)$ that the average number of balls  in one urn over time $T$,  $\bar{n}$, takes any specified value $aN$, where $0\leq a\leq 1$.  For long observation time, $T\to \infty$, a Donsker-Varadhan large deviation principle  holds: $-\ln \mathcal{P}_T(\bar{n}= a N) \simeq T I(a,N,K,\dots)$, where
$\dots$ denote additional parameters of the model. We calculate the rate function $I(a,N,K, \dots)$ exactly by two different methods due to Donsker and Varadhan and compare the exact results with those obtained with a variant of WKB  approximation (after  Wentzel, Kramers and Brillouin). In the absence of interactions the WKB prediction for $I(a,N,K, \dots)$ is exact for any $N$. In the presence of interactions the WKB method gives asymptotically exact results for $N\gg 1$. The WKB method also uncovers the (very simple) time history of the system which dominates the contribution of different time histories to $\mathcal{P}_T(\bar{n}= a N)$.

\end{abstract}
\maketitle

\tableofcontents
\nopagebreak

\section{Introduction}
\label{intro}

The Ehrenfest urn model (EUM) \cite{EUM} was referred to by Marc Kac
as ``probably one of the most instructive models in the whole of physics" \cite{Kac}. In the basic version of the EUM there are two urns with the total of $N$
particles. At each time step a particle is chosen randomly and moved to the other urn. The original
goal of the EUM was to understand how thermodynamic irreversibility emerges from time-reversible dynamics. This model was studied by Kohlrausch and Schr\"{o}dinger \cite{Schroedinger}, and the original questions about it were  fully answered by Kac \cite{Kac1}, Siegert \cite{Siegert} and
Hess \cite{Hess}. The equilibrium distribution of the number of balls in one of the urns is binomial. The characteristic time to reach this equilibrium scales as $N$. In contrast,  the Poincar\'{e} recurrence time (the average time it takes the system to visit the  state farthest from the equilibrium) scales as $2^N$ and becomes prohibitively large at large $N$ thus explaining thermodynamic irreversibility \cite{Kac}.

As it often happens with simple but rich and insightful models, the EUM was generalized in many ways and found applications in many areas of science, from reliability theory \cite{Keilson,Takacs} to genetics \cite{Karlin}. In particular, it continues to help elucidate the mechanisms of ``disparity between the time symmetry of the fundamental laws of physics and the time asymmetries of the observed universe" \cite{Gell-Mann}. Here we will consider this model in the context of large deviations of \emph{time-additive} quantities  which have attracted recent interest in statistical mechanics \cite{Touchette2009,Touchette}.

Suppose that we observe a realization of the stochastic dynamics of the EUM. We count the number of balls $n(t)$ in one of the urns (let us call it urn 1) as a function of time and calculate the empirical average of this number over time $T$:
\begin{equation} \label{nbarDefinition}
\bar{n} = \frac{1}{T}\int_{0}^{T}n(t)dt ,
\end{equation}
where $n(t) \in \{0,1, \dots, N\}$. Let $T$ go to infinity. What is the probability, $\mathcal{P}_T(\bar{n}= a N)$, that $\bar{n}$ is equal to a specified number $aN$, where $0\leq a\leq 1$? For the two-urn model, the \emph{expected} value of $\bar{n}$ is equal to $N/2$. The probability of any deviation from $N/2$ is exponentially small in $T$ and encoded in the Donsker-Varadhan large-deviation principle \cite{DonskerVaradhan}:
\begin{equation}\label{DVscaling}
-\ln \mathcal{P}_T(\bar{n}= a N) \simeq T I(a,N),
\end{equation}
where the rate function $I(a,N)$ is a convex function of $a$ which vanishes at $a=1/2$.
Our primary goal will be to calculate $I(a,N)$ for the EUM and for some of its extensions.  We will be interested in the continuous-time version of the model \cite{Karlin}, where transition time is a random quantity, distributed with a negative exponential distribution. We choose the mean time of this distribution to be $1$.

An immediate generalization is to $K$ fully-connected urns, so that the rate function $I(a,N,K)$ also depends on $K$.  In the standard version of the EUM the balls are non-interacting: the transition rate of a ball to another urn is constant. In the absence of interactions each of the balls behaves independently from the other balls. As a result, the probability of a time-additive quantity is a product of the corresponding single-ball probabilities, and the rate function $I(a,N,K)$ must be proportional to $N$:
\begin{equation}\label{Nscaling}
I(a,N,K)= N f(a,K).
\end{equation}
When the balls interact, the transition rate of a ball to another urn depends on the current number of balls in the departure urn and/or target urn. In this case this simple scaling of $I(a,N,K)$ with $N$ breaks down.

Here we consider both generalizations ($K>2$ and interactions) and use the EUM to compare
three different  methods for calculating the  rate function $I(a,N,K)$. Two of them are exact. The most general first method was developed by Donsker and Varadhan (DV) \cite{DonskerVaradhan} , and it employs the G\"{a}rtner-Ellis theorem \cite{GartnerEllis}. The second method, also due to DV \cite{DonskerVaradhan}, relies on the absence of interactions among balls and on the equilibrium properties of the model.
The third method is a variant of the  WKB approximation \cite{Kubo,Gang,Dykman,EK,AssafMeerson}. It is also called in some circles  the optimal fluctuation method, or the weak-noise theory. The WKB method requires a large parameter $N\gg 1$. In this regime we obtain an approximate result
\begin{equation}\label{Nscalingnonlinear}
I(a,N,K)\simeq g(N) f(a,K),\quad \quad N\gg1,
\end{equation}
where the function $g(N)$ is, in general, nonlinear.

As we show here, the WKB method is simple and robust. It is  especially useful in situations where the first DV method can be implemented only numerically, whereas the second DV method is inapplicable because of interactions. WKB approximation for continuous stochastic processes, known as the Freidlin-Wentzell theory \cite{FM}, has been already
applied to large deviations of time-additive  quantities  \cite{Seifert,Engel,TouchetteWKB}.  In contrast, WKB approximation for jump processes \cite{Kubo,Gang,Dykman,EK,AssafMeerson}  has not yet found applications to this type of large deviations. Our work closes this gap.

\section{Donsker-Varadhan Methods}
\label{exact}

First, we briefly describe the general DV  method \cite{DonskerVaradhan}
(see Ref. \cite{Touchette} for an accessible exposition), which allows one to determine the rate function exactly, but rarely analytically. The continuous-time EUM is governed by a master equation which describes the evolution of the probability $P_n(t)$ of observing $n$ balls at time $t$ in urn 1:
\begin{equation} \label{masterGeneral}
\frac{d}{dt} P_n(t) = \hat{L}\,P_n(t).
\end{equation}
The explicit form of the matrix operator $\hat{L}$ will be presented shortly. We condition this Markov process by the relation
$\bar{n}=aN$, where $\bar{n}$ is defined by Eq.~(\ref{nbarDefinition}). The general DV method involves the calculation of the rate function $I(a,N)$ via the Legendre-Fenchel transform,
\begin{equation} \label{legendre}
I(a,N) = \max_{k} [k\bar{n} - \lambda(k,N)],
\end{equation}
of the scaled cumulant generating function $\lambda(k,N)$, introduced by G\"{a}rtner and Ellis \cite{GartnerEllis}:
\begin{equation} \label{lambda}
\lambda(k,N) = 
\lim_{T \to \infty}\frac{1}{T}\ln\langle e^{Tk\bar{n}(T)}\rangle\,,
\end{equation}
where $\langle \dots\rangle$ denotes  averaging over all possible values of $\bar{n}$\footnote{In general, one needs to specify the initial condition $n(t=0)$, 
but in the long-time limit, $T \to \infty$, it has no effect if the process is ergodic.}.  The  DV method maps the problem of finding the large deviation function of a long-time average into a spectral problem. According to this method, $\lambda(k,N)$ is given by the maximum eigenvalue of an auxiliary matrix operator $\hat{L}^k$:
\begin{equation} \label{donskerVaradhan}
\lambda(k,N) = \xi_{\text{max}}(\hat{L}^k),
\end{equation}
where
\begin{equation} \label{tiltedConjOperator}
L^k_{n,m} = (L^+)_{n,m} + nk\delta_{n,m} ,
\end{equation}
which is a tilted version of the Hermitian conjugate of $\hat{L}$.

\begin{figure}[h]
\includegraphics[width=0.4\textwidth,clip=]{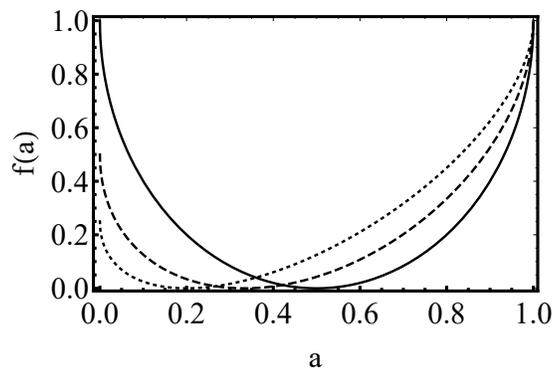}
\caption{The rescaled large-deviation function $f = I/N$ vs. $a$ for $K = 2$ (solid), $K = 3$ (dashed) and $K = 5$ (dotted).}
\label{ratesKurns}
\end{figure}

Now we specify this formalism to the EUM with any number of urns $K>1$. Here the general DV method can be implemented analytically.  When the balls do not interact, their arrangement in the rest of urns do not affect their transition rates to urn 1. This brings us to an effective two-urn problem. There are urn 1, on which we focus, and a ``superurn" which represents all the other urns combined. The superurn can have inner transitions, but they do not affect the state of the first urn. Therefore, the master equation is
\begin{equation} \label{masterKurns}
\partial_t P_n(t) = \frac{N-n+1}{K-1}P_{n-1}(t) + (n+1)P_{n+1}(t) - \left(n + \frac{N-n}{K-1}\right)P_n(t)
\end{equation}
According to Eqs.~(\ref{masterGeneral}) and (\ref{tiltedConjOperator}), the tilted operator $\hat{L}^k$ is given by
\begin{equation} \label{tiltedKurns}
\hat{L}^k = n \delta_{n-1, m} + \frac{N-n}{K-1}\delta_{n+1, m} + \left[nk - \left(n + \frac{N-n}{K-1}\right)\right]\delta_{n,m} .
\end{equation}
The maximum eigenvalue of this operator  is\footnote{Since, for the non-interacting particles,  the rate function must be proportional to $N$, it suffices to calculate the maximum eigenvalue for $N=1$}
\begin{equation} \label{lambdaKurns}
\lambda(k, N, K) =
\frac{N}{2}\left(k - \frac{K - \sqrt{(k-1)^2K^2 - 2k(k-3)K + k(k-4)}}{K - 1}\right) .
\end{equation}
Substituting this result into Eq.~(\ref{legendre}), we find the rate function in the form of Eq.~(\ref{Nscaling}), where
\begin{equation}\label{rateKurns}
f(a, K) =  \left(\sqrt{a}-\sqrt{\frac{1-a}{K-1}}\right)^2 .
\end{equation}
This function, which is strictly convex, is shown in Fig.~\ref{ratesKurns}.  $f(a,K)$ has its minimum, and vanishes, at the expected value $\bar{n}=N/K$, that is  $a=1/K$.
For $K=2$ the rate function is symmetric around $a=1/2$. For larger values of $K$, the balls are more likely to be in the other urns, so the minimum of $f(a,K)$ is shifted towards smaller $a$. The maximum value of $f$ is reached at $a=1$: $f(1,K)=1$. The corresponding probability $\mathcal{P}_T =\exp(-TN)$ describes a  rare event when neither of the $N$ balls, located in urn $1$, is chosen during time $T$, or a fraction of this time which tends to $1$ as $T$ goes to infinity. (To remind the reader, in our continuous-time Markov process the transition times are exponentially distributed with average $1$.) Another extreme case, $f(0,K)=1/(K-1)$ describes the situation when neither of the $N$ balls, located in the urns $2, \dots, K$, hops to urn $1$ during time $T$.
The factor $1/(K-1)$  describes the fraction of hops into 
urn $1$ among all possible hops in the system.

As we will see shortly, the total number of balls $N$ factors out from the rate function also for interacting balls, but only in the large-$N$ limit. Before that, however, we
rederive the exact rate function (\ref{rateKurns}) by using the second DV method \cite{DonskerVaradhan}. This method deals with what is known in the mathematical literature as ``level 2 large deviations"  \cite{Touchette2009}, where one is interested in large deviations of the empirical measure $\rho_T(n)$, where $n=0,1, \dots $ numbers the discrete states of the system. For the EUM with $N=1$ the empirical measure is defined as
\begin{equation}\label{empirical}
\rho_T(n) = \frac{1}{T} \int_0^T \delta (n_t -n) dt, \quad n=0,1 ,
\end{equation}
which is the fraction of time the ball spends in the superurn
and in urn 1, respectively.  The level-2 method assumes detailed balance, which the EUM satisfies. In addition, it relies on the knowledge of the equilibrium probability distribution. The latter, for the EUM with $N=1$ ball, is simply $\pi_0=1-1/K$ and $\pi_1=1/K$. According to DV \cite{DonskerVaradhan}, the rate function of the large deviations of the empirical measure is equal to
\begin{equation}\label{level2}
- \mathlarger \sum_{n=0}^1 \,\mathlarger \sum_{m=0}^1 \pi_n \sqrt{\frac{\rho_T(n)}{\pi_n}} \,L_{n,m} \,\sqrt{\frac{\rho_T(m)}{\pi_m}}=\left(\sqrt{
  \rho_T(1)}-\sqrt{\frac{\rho_T(0)}{K-1}}\right)^2.
\end{equation}
Setting $\rho_T(1)=a$ and, by virtue of the probability conservation, $\rho_T(0)=1-a$,  we immediately arrive at Eq.~(\ref{rateKurns}).

\section{WKB Method}
\label{wkbmethod}
Before we introduce the WKB method, let us extend the EUM by allowing
interactions between the balls in the departure urn. Specifically, we will consider a variant of the zero range process \cite{evans}, where the transition rate of a ball from one urn to another is a power-law function, with exponent $\alpha>0$, of the number of balls in the departure urn\footnote{A more complicated interaction between the balls in the same urn has been recently considered in Ref. \cite{taiwan}, where the relaxation time to the equilibrium state, the Poincar\'{e} cycles and phase transitions have been studied.}.

For interacting balls in $K$ urns, the master equation cannot be written in terms of an urn and a superurn, as we did for the case of non-interacting balls. Therefore, we must deal with the evolution of the multivariate probability $P_{n_1,n_2,\dots , n_{K-1}}(t)$ of observing $n_1$ balls in the first urn, $n_2$ balls in the second urn, $\dots$, and $n_{K-1}$ balls in the $(K-1)$-th urn.  The population of the $K$-th urn is set by the conservation of the total number of balls. For example, for $K=3$ urns
the master equation takes the form
\begin{eqnarray}
  \frac{d}{dt}P_{n_1,n_2} (t) &=& \frac{1}{2} (N-n_1-n_2+1)^{\alpha} P_{n_1-1,n_2}(t)+\frac{1}{2} (n_1+1)^{\alpha} P_{n_1+1,n_2} (t)\nonumber\\
  &+&\frac{1}{2} (N-n_1-n_2+1)^{\alpha} P_{n_1,n_2-1}(t)+ \frac{1}{2} (n_2+1)^{\alpha} P_{n_1,n_2+1} (t) \nonumber\\
  &+& \frac{1}{2} (n_1+1)^{\alpha} P_{n_1+1,n_2-1} (t) +\frac{1}{2} (n_2+1)^{\alpha} P_{n_1-1,n_2+1}(t)\nonumber\\
  &-&  \left[ (N-n_1-n_2)^{\alpha} +n_1^{\alpha}+n_2^{\alpha}\right]P_{n_1,n_2} (t). \label{master3alpha}
\end{eqnarray}
In the absence of interactions one has $\alpha=1$.

When $\alpha \neq 1$, the level-2 DV method is inapplicable, whereas the main DV method can be implemented only numerically.
The WKB method gives a viable analytic alternative when $N\gg 1$.
The method employs this large parameter in a smart way. For jump processes  -- Markov processes with continuous time and discrete space, such as the one described by Eq.~(\ref{master3alpha}) -- the WKB method was put forward by Kubo, Matsuo and Kitahara \cite{Kubo}, see also Refs. \cite{Gang,Dykman}. In the last decade the method has been developed much further and used in different applications, mostly in the context of stochastic reactions and stochastic population dynamics, see Ref. \cite{AssafMeerson} for a recent review.
The WKB method  approximately solves the master equation via an exponential ansatz for $P_{n_1,n_2, \dots, n_{K-1}} (t)$. For our zero-range process the exponential ansatz is
\begin{equation}\label{WNTansatz}
P_{n_1,n_2, \dots, n_{K-1}} (t) = e^{-N^{\alpha} s\left(\frac{n_1}{N},\frac{n_2}{N},\dots, \frac{n_{K-1}}{N},t \right)}, \quad N\gg 1 .
\end{equation}

Now we treat $q_i=n_i/N$ as continuous variables and Taylor-expand the function $s$ of shifted arguments, such as in
\begin{eqnarray}
 && s\left(\frac{n_1-1}{N},\frac{n_2}{N},\dots, \frac{n_{K-1}}{N},t \right)\equiv  s\left(q_1-\frac{1}{N},q_2,\dots, q_{K-1},t \right) \nonumber\\
  &&=s\left(q_1,q_2,\dots, q_{K-1},t \right) -
\frac{1}{N}\,\frac{\partial s\left(q_1,q_2,\dots, q_{K-1},t \right)}{\partial q_1}+\frac{1}{2N^2} \frac{\partial^2 s\left(q_1,q_2,\dots, q_{K-1},t \right)}{\partial q_1^2} + \dots.
\end{eqnarray}
In the leading order in $1/N$ we omit the second- and higher-order terms and arrive at a nonlinear first-order PDE for $s(q_1,q_2,\dots q_{K-1},t)$ which has the form of a Hamilton-Jacobi equation for the action $s$,
\begin{equation}\label{HJ}
\frac{\partial s}{\partial t} + H_0\left(q_1,q_2,\dots , q_{K-1}, \frac{\partial s}{\partial q_1},\frac{\partial s}{\partial q_2}, \dots , \frac{\partial s}{\partial q_{K-1}},\, t\right)=0,
\end{equation}
with a problem-specific Hamiltonian $H_0$. In our example, Eq.~(\ref{master3alpha}), $H_0$ does not depend explicitly on time and can be written as
\begin{equation}\label{H03}
 H_0(q_1,q_2,p_1,p_2) = \frac{1}{2}\left(1-q_1-q_2\right)^{\alpha} \left(e^{p_1}+e^{p_2}-2\right)+\frac{1}{2} q_1^{\alpha} \left(e^{-p_1+p_2}+e^{-p_1}-2\right)
+ \frac{1}{2} q_2^{\alpha} \left(e^{p_1-p_2}+e^{-p_2}-2\right) .
\end{equation}
The ensuing Hamilton equations should be solved subject to the proper boundary conditions in the time domain, see below.

Once the coordinates $q_1(t)$ and $q_2(t)$
and the momenta $p_1(t)$ and $p_2(t)$ are found,  the effective action $s(q_1,q_2,t)$ can be determined from the equation
\begin{equation}\label{action}
s(q_1,q_2,t) = \int (p_1 dq_1+p_2 dq_2 -H_0 dt) =
\int_0^t \left[p_1(t) \dot{q}_1(t) +p_2(t) \dot{q}_2(t)\right] dt-E t,
\end{equation}
where
\begin{equation}\label{energy}
E=H_0[q_1(t),q_2(t),p_1(t),p_2(t)]
\end{equation}
is the (conserved) energy of the Hamiltonian trajectory.

The Hamiltonian (\ref{H03}) describes unconstrained dynamics.
The invariant plane $p_1=p_2=0$ corresponds to the deterministic (zero-noise) theory, where $q_1$ and $q_2$ flow into the attracting fixed point $(1/3,1/3)$ which, in the original variables,  describes equal numbers of balls, $N/3$ in each of the three urns. 
In the 4-dimensional phase space of the Hamiltonian flow this fixed point becomes $M_0=(1/3,1/3,0,0)$. This fixed point is a saddle point: in addition to the two stable manifolds, confined to the plane $p_1=p_2=0$, it has two unstable manifolds which involve $p_1\neq 0$ and $p_2\neq 0$. This fixed point is the only fixed point of this simple Hamiltonian flow.

Conditioning the original stochastic process on a specified time average of $n_1$ implies constraining the Hamiltonian flow
by the equation
\begin{equation}\label{averageq}
\frac{1}{T} \int_0^T q_1(t) dt = a.
\end{equation}
This constraint can be accounted for via a Lagrange multiplier $\Lambda$. In the Hamiltonian description this leads to
the modified Hamiltonian
\begin{equation}\label{Hlambda}
H(q_1,q_2,p_1,p_2;\Lambda) = H_0(q_1,q_2,p_1,p_2)+\Lambda q_1.
\end{equation}
As the unconstrained flow, the constrained flow has a single fixed point $M$ which is a saddle point. It is more convenient to parametrize this fixed point
by its value of $q_1\equiv b$, rather than by the  Lagrange multiplier  $\Lambda$:
\begin{equation}\label{M}
q_1=b,\quad q_2=\frac{1-b}{2},\quad p_1=\frac{\alpha}{2}  \ln \left(\frac{2b}{1-b}\right), \quad p_2=0 .
\end{equation}
Notice that, for any $0\leq b\leq 1$, the numbers of balls in the second and third urn, as described by the fixed point $M$, are equal to each other. This equidistribution is to be expected from a minimum-action solution. The particular case of $b=1/3$ corresponds to the fixed point of the unconstrained system, $\Lambda=0$.

For a specified $T$ the proper Hamiltonian trajectory (called the activation trajectory) is determined by the initial and final conditions, such as $q(t=0)=q_{\text{in}}$ and  $q(t=T)=q_{\text{fin}}$. The resulting action (\ref{action}) must be minimized with respect to $q_{\text{fin}}$. For arbitrary $T$ the calculations are tedious.
At very large $T$, however, it becomes very simple. This is because, as $T$ increases, the ``particle" following the activation trajectory
spends a progressively longer time in a close vicinity of the fixed point $M$. As $T\to \infty$, the dominant contribution to the action (\ref{action}) comes from the fixed point $M$ itself, where we should set $b=a$ by virtue of the condition Eq.~(\ref{averageq}). In this limit we obtain $q(t=T)\to a$, whereas the action becomes independent of the initial condition $q(t=0)$. Furthermore, each of the two terms under the integral over $t$ in Eq.~(\ref{action}) vanishes, and we obtain
\begin{equation}\label{s3}
s=-E T = -T H_0\left[a,\frac{1-a}{2},\frac{\alpha}{2}  \ln \left(\frac{2a}{1-a}\right),0\right]= T\left[a^{\alpha/2}-\left(\frac{1-a}{2}\right)^{\alpha/2}\right]^2.
\end{equation}
The resulting large-$N$ asymptotic rate function is $I(a,N,K=3,\alpha)=N^{\alpha} f(a,K=3,\alpha)$, where
\begin{equation}\label{rf3}
f(a,K=3,\alpha)= \left[a^{\alpha/2}-\left(\frac{1-a}{2}\right)^{\alpha/2}\right]^2 .
\end{equation}
For an arbitrary number $K>1$  of urns the calculations are almost identical. Here too, at $T\to \infty$, the dominant contribution to the action, and therefore to the rate function,  comes from a unique saddle point of the constrained Hamiltonian, with coordinates
\begin{equation}
  q_1 = a, \quad  p_1=\frac{1}{2} \ln \frac{(K-1)a}{1-a}, \quad
  q_i = \frac{1-a}{K-1},\quad p_i=0, \quad i=2,3,\dots K-1 . \label{Mk}
\end{equation}
Again, for any specified $a$ in the first urn, the numbers of balls in each of the other urns are equal to each other.  The resulting rate function for $K$ urns is
\begin{equation}
\label{rfk}
  I(a,N,K,\alpha)\simeq N^{\alpha} f(a,K,\alpha), \quad f(a,K,\alpha) = \left[a^{\alpha/2}-\left(\frac{1-a}{K-1}\right)^{\alpha/2}\right]^2.
\end{equation}

\begin{figure}[h]
\includegraphics[width=0.4\textwidth,clip=]{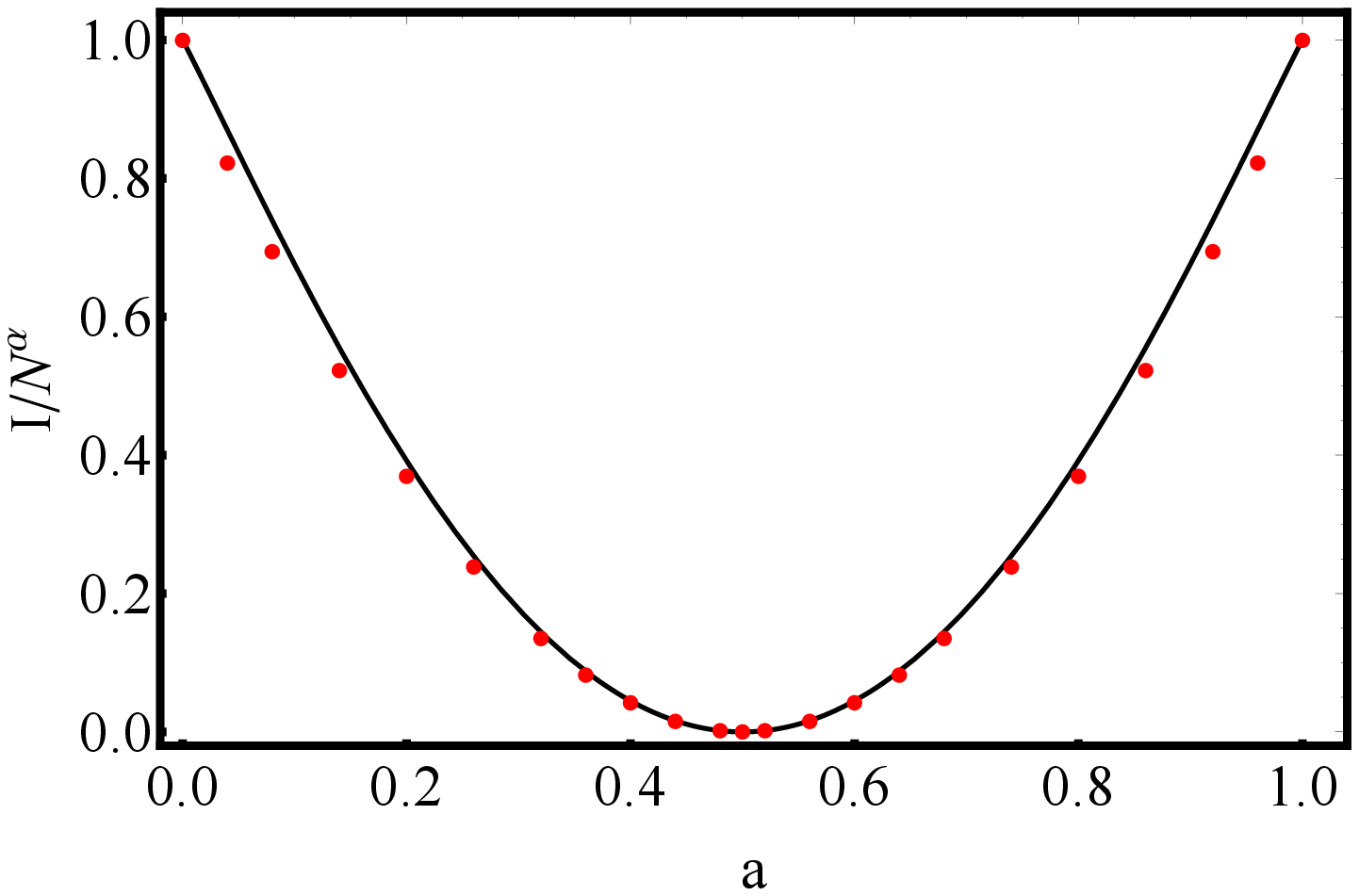}
\includegraphics[width=0.4\textwidth,clip=]{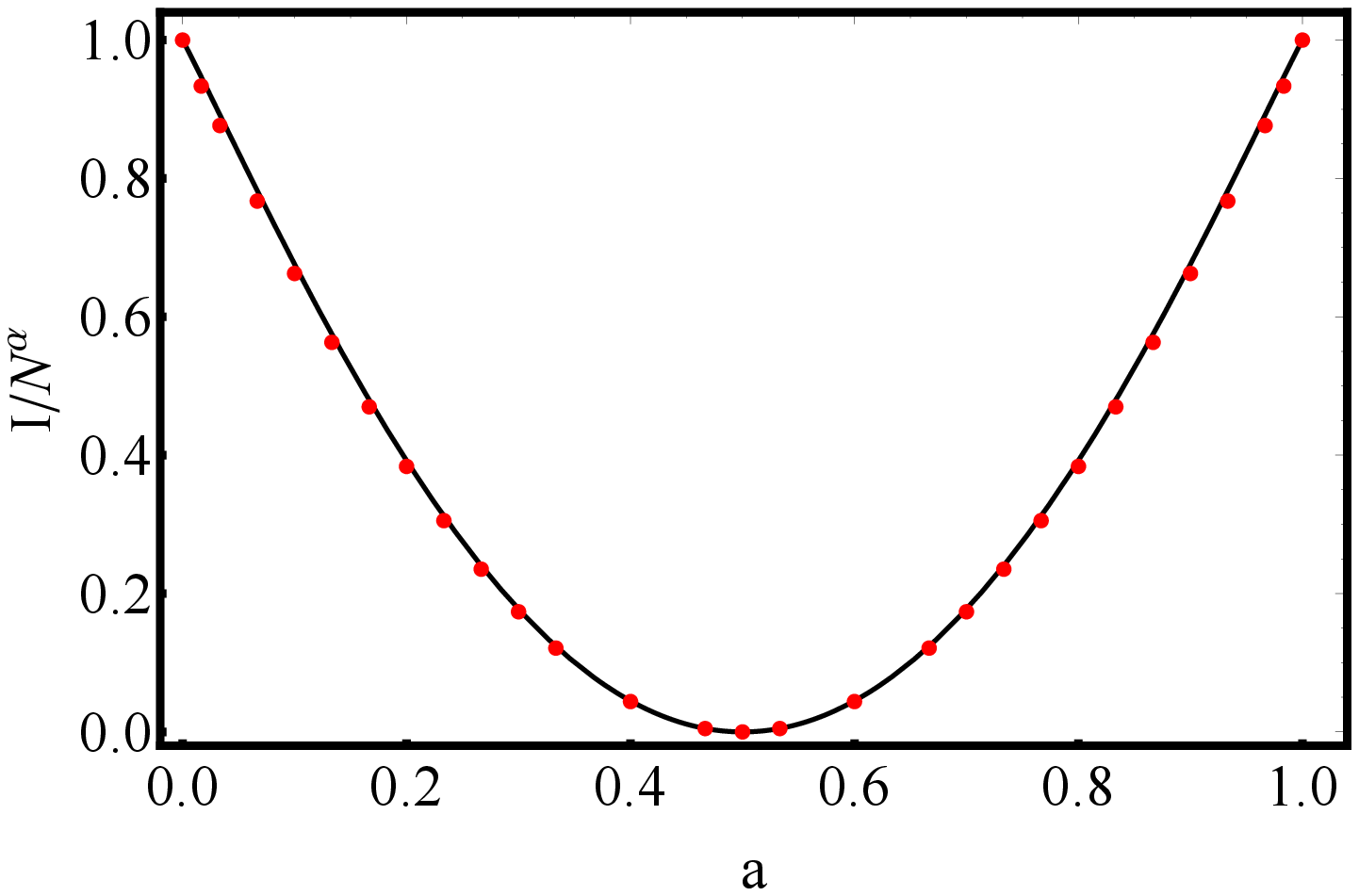}
\caption{The rescaled rate function $I/N^{\alpha}$ vs. $a$ for $\alpha=3$ and $K=2$, as predicted by the function $f$ from Eq.~(\ref{rfk}) (solid line), is compared with accurate numerical results with the main DV method (symbols) for $N=5$ (left panel) and $N=15$ (right panel).}
\label{comparison}
\end{figure}

We shall now discuss the obtained result \eqref{rfk}.
In the absence of interactions, $\alpha=1$, Eq.~(\ref{rfk}) coincides with the exact expression (\ref{rateKurns}).  For $\alpha\neq 1$ Eq.~(\ref{rfk}) is an approximation, valid only at $N\gg 1$. For small $N$ it does not apply. For example, when there is only one particle, $N=1$, there are no interactions, and the exact $f(a,K)$  must coincide with Eq.~(\ref{rateKurns}) and be independent of  $\alpha$, in contradiction with Eq.~(\ref{rfk}). At large $N$, however, the accuracy of approximation Eq.~(\ref{rfk}) is good and improves with an increase of $N$.  Notice that, for $a=1$, the WKB result (\ref{rfk}), where $f(1,K,\alpha)=1$, is exact for any $N$. It describes the rare event when all $N$ balls, located in urn $1$, remain there for the whole time $T$ or a fraction of this time which goes to $1$ as $T\to\infty$. For $K=2$ one has $f(a,2,\alpha)=f(1-a,2,\alpha)$, which is to be expected from symmetry arguments. Figure~\ref{comparison} compares, for $\alpha=3$, the WKB rate function (\ref{rfk}) with that found numerically with the main DV method. To avoid cumbersome numerics, we made the comparison for $K=2$.

\begin{figure}[h]
\includegraphics[width=0.4\textwidth,clip=]{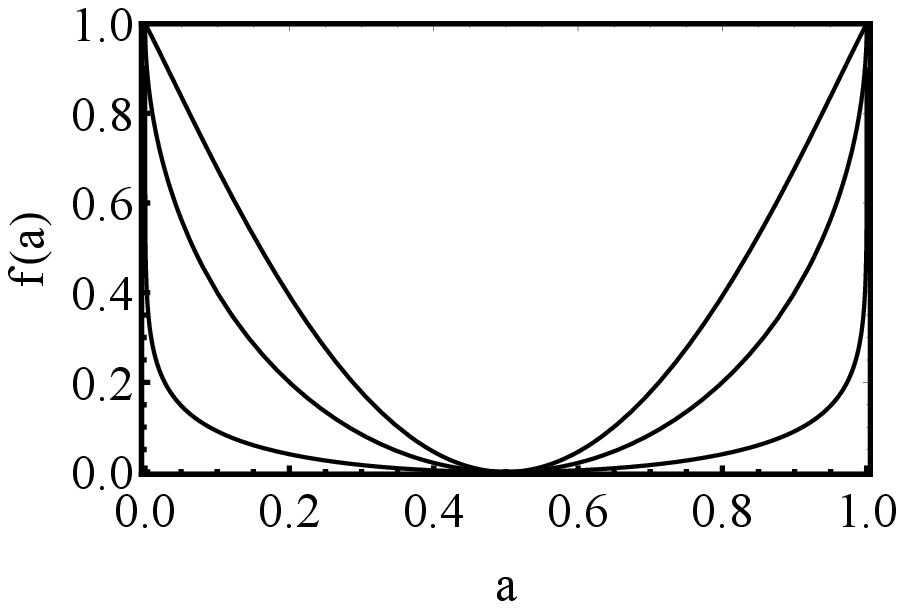}
\includegraphics[width=0.4\textwidth,clip=]{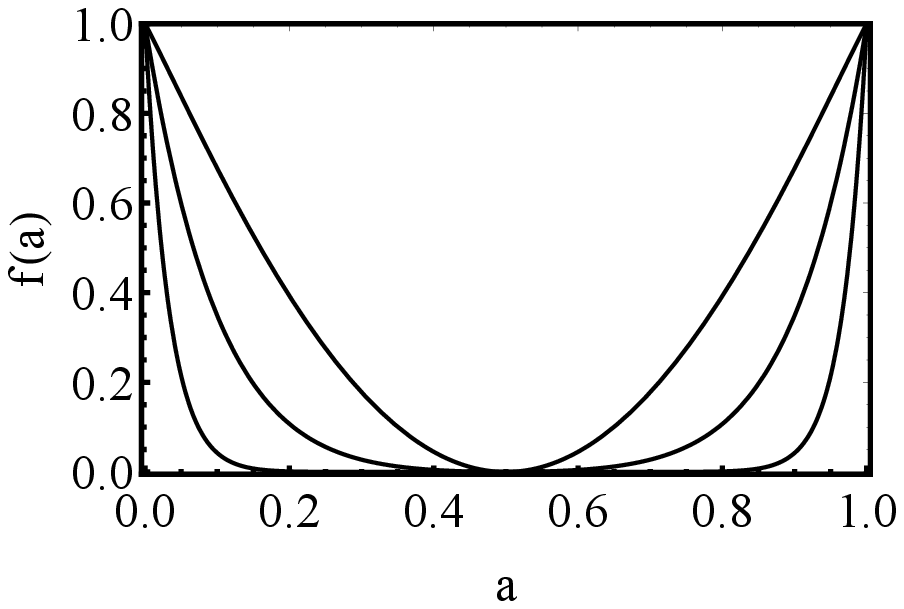}
\caption{The rescaled rate function $f=I/N^{\alpha}$ vs. $a$ for $K=2$ and different $\alpha$. Left panel,
bottom to top: $\alpha=1/3$, $1$ and $2/\ln 2=2.8853\dots$. Right panel, top to bottom: $\alpha=2/\ln 2$, $10$ and $30$.}
\label{falpha}
\end{figure}

\begin{figure}[h]
\includegraphics[width=0.4\textwidth,clip=]{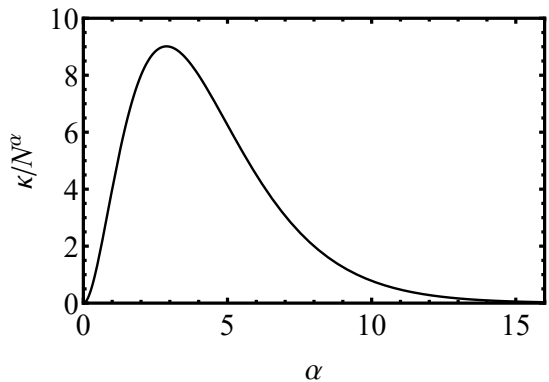}
\caption{The rescaled curvature of the rate function $N^{-\alpha} \kappa(\alpha, K=2) = 2^{3-\alpha}\alpha^2$ vs. $\alpha$ for $K=2$. The maximum is at $\alpha=2/\ln 2$.}
\label{curvature}
\end{figure}

As $\alpha$ tends to zero or to infinity, the function $f$ flattens as a function of $a$, except at $a$ close to $0$ and $1$. This can be seen on Fig. \ref{falpha} which shows $f$ versus $a$ for $K=2$ urns and several values of $\alpha$. This effect can be characterized by
the rescaled curvature $N^{-\alpha} \kappa(\alpha, K)$
of the rate function (\ref{rfk})  at its minimum point $a=1/K$. This quantity is inversely proportional to the variance of the distribution $\mathcal{P}_T(a)$:
\begin{equation}\label{curvK}
\frac{\kappa(\alpha, K)}{N^{\alpha}} \equiv \frac{\partial^2 f}{\partial a^2}\Big|_{a=1/K}= \frac{\alpha^2 K^{4-\alpha }}{2 (K-1)^2} .
\end{equation}
It tends to zero at $\alpha \to 0$ and $\alpha \to \infty$
and reaches its maximum  at $\alpha=2/\ln K$, see Fig.  \ref{curvature}.   At present we do not have a good intuitive explanation for this somewhat surprising non-monotonic behavior.

The rescaled curvature (\ref{curvK}) can be also obtained from  still another approximate method, when the exact master equation (\ref{master3alpha}) is approximated by the Fokker-Planck equation via the van Kampen system-size expansion \cite{Kac1,Kampen}. In this way the EUM, conditioned by Eq. (\ref{nbarDefinition}), is approximated by the Ornstein-Uhlenbeck process, conditioned in the same way. The rate function of the latter (a parabolic function of $a$ around $a=1/K$) can be easily calculated \cite{Touchette}. This rate function, however, describes correctly only the Gaussian asymptotic of the distribution $\mathcal{P}_T$ in a close vicinity of $\bar{n}= N/K$.

Going back to Eq.~\eqref{rfk}, we observe that it does have a structure of an effective system of an urn and a superurn, just as in the non-interacting case. This is true, however, only in the WKB approximation, and only because of the fact that  Eq.~\eqref{rfk}  comes from the fixed point \eqref{Mk} which is symmetric with respect to the \emph{unconditioned} urns.

\section{Discussion}
\label{summary}
We evaluated the large deviation function of the long-time-average number of balls in one of the urns of the EUM.
We also tested the validity of the WKB method for evaluating the rate function $I(a,N,K, \dots)$ by comparing the WKB predictions with the exact results from two different methods due to Donsker and Varadhan. In the absence of interactions between the balls the WKB method yields exact result for $I(a,N,K)$ for any $N$. In the presence of interactions, modeled here as a zero-range process, the WKB method gives asymptotically exact results for $N\to \infty$. The WKB method also uncovers the (very simple) time history of the system, which dominates the contribution of different time histories to $\mathcal{P}_T(\bar{n}= a N)$.

The main DV method requires that the averaging time $T$ be large, but $N$ can be arbitrary. The WKB method requires $N\gg 1$ but, in general, is not limited to large $T$. In this sense the main DV method and the WKB method can be considered as complementary.

One natural extension of the EUM would be spatially explicit: the urns are connected \emph{not} via a complete graph. We argue, on intuitive grounds,  that the rate function, that we evaluated here for fully-connected urns, will serve as an upper bound for the corresponding  less-connected case. A simple example of a less-connected system is a one-dimensional lattice of $K$ urns with periodic boundary conditions, where balls can hop only between neighboring urns. An analysis of this problem (which confirms the above-mentioned bound) will be reported elsewhere.

It would be also interesting to extend the WKB method of evaluating large-deviation functions of time-additive quantities
to jump processes which deterministic description predicts multiple fixed points \cite{AssafMeerson}.

\section*{Acknowledgments}
We are very grateful to Hugo Touchette for advice and a critical reading of the manuscript. This research was supported by the Israel Science Foundation (Grant No. 807/16).

\end{document}